\documentclass[preprint,prd,showpacs,preprintnumbers,amsmath,amssymb]{revtex4}

\usepackage{graphicx}
\usepackage{dcolumn}
\usepackage{bm}
\usepackage{amsfonts}

\newcommand{\nl}{\nonumber \\ }

\newcommand{\order}{ {\cal O}}
\newcommand{\be}{\begin{equation}}  
\newcommand{\ee}{\end{equation}}  
\newcommand{\bear}{\begin{eqnarray}}  
\newcommand{\eear}{\end{eqnarray}}  
\newcommand{\ba}{\begin{array}}  
\newcommand{\ea}{\end{array}}

\newcommand{\tp}{\tilde{\pi}} 
 
\newskip\humongous \humongous=0pt plus 1000pt minus 1000pt

\newif\ifdtup

\def\oldreffmt#1{\rlap{[#1]} \hbox to 2\parindent{}}

\def\figfmt#1{\rlap{Figure {#1}} \hbox to 1in{}}  
  
\def\ie{\hbox{\it i.e.}{}}      
  
\def\eg{\hbox{\it e.g.}{}}

\def\Tr{\mathop{\rm Tr}}  
\def\Im{\mathop{\rm Im}}  
\def\Re{\mathop{\rm Re}}

\def\VEV#1{\left\langle #1\right\rangle}

\def\slash#1{#1\!\!\!/\!\,\,}  
\def\beq{\begin{equation}}  
\def\eeq{\end{equation}}  
\def\bea{\begin{eqnarray}}  
\def\eea{\end{eqnarray}}  
\def\half{\frac{1}{2}}  
  
\def\bq{\begin{quote}}  
\def\eq{\end{quote}}

\def\half{\frac{1}{2}}       

\relax  

\newdimen\tdim  
\tdim=\unitlength  
\def\mybar#1{\overline{\!#1\!}}

\begin{document}  

\preprint{FERMILAB-PUB-07/004-T} \title{Topological Physics of Little
  Higgs Bosons }

\author{Christopher T. Hill \\ and \\
  Richard J. Hill}

{\email{hill@fnal.gov}}

\affiliation{
  {{Fermi National Accelerator Laboratory}}\\
  {{\it P.O. Box 500, Batavia, Illinois 60510, USA}} }

\date{5 January, 2007} 

\begin{abstract}  
  Topological interactions will generally occur in composite Higgs or
  Little Higgs theories, extra-dimensional gauge theories in which
  $A_5$ plays the role of a Higgs boson, and amongst the pNGB's of
  technicolor.  This phenomena arises from the chiral and anomaly structure of
  the underlying UV completion theory, and/or through chiral
  delocalization in higher dimensions.  These effects are described by
  a full Wess-Zumino-Witten term involving gauge fields and pNGB's. We
  give a general discussion of these interactions, some of which may
  have novel signatures at future colliders, such as the LHC and ILC.
\end{abstract}



\maketitle

\section{\bf Introduction}

There is a fundamental topological distinction between an ordinary
Higgs boson and a composite \cite{Kaplan:1983fs} or Little Higgs boson 
\cite{nima}.  
The ordinary Higgs boson is
described by a scalar field taking values on a flat, unbounded
manifold.  The composite or Little Higgs boson, on the other hand, is a (pseudo)
Nambu-Goldstone boson (pNGB) of a spontaneously broken chiral
symmetry, the analog of a meson such as the kaon (we'll henceforth
refer to all such composite Higgs bosons as Little Higgs bosons).  
It is described
by a field that is confined to a compact manifold, of ``radius''
$1/F$.  The Little Higgs is thus an angular variable and it can be
translated through $2\pi F$ circumnavigating the manifold.  As a
consequence, topologically stable configurations can exist that are
described by conserved topological currents.  These currents arise
from the full effective action only if we include a new topological
interaction, known as the Wess-Zumino-Witten (WZW)
term~\cite{Wess:1971yu,Witten}.  This term is the low-energy effective
description of anomaly physics in terms of pNGB's and gauge
fields~\cite{BJ,Adler,bardeen}.  It unifies the topological physics in
a chiral Lagrangian, and describes new physical processes that are not
expected for the ordinary Higgs boson.  It is tied to any particular
UV completion model through integer quantities such as the number of
``techni-colors'' of the constituent ``techni-quarks.''  Thus, the new
WZW interactions of Little Higgs bosons probe the underlying UV
completion theory, much like the $\pi^0\rightarrow \gamma\gamma$
interaction probes QCD.

To motivate our discussion, we recall the typical interactions
contained in the ungauged Wess-Zumino~\cite{Wess:1971yu} term of QCD,
\begin{eqnarray} 
\label{onee}
\int \Tr( \tp d\tp d\tp d\tp d\tp )
&=& 
\int\! d^4x\, \bigg[ 5 K^\dagger \partial_\mu\pi \partial_\nu\pi \partial_\rho\pi 
\partial_\sigma K 
-{5\sqrt{3}} \eta\; \partial_\mu K^\dagger \partial_\nu K
\partial_\rho K^\dagger \partial_\sigma K  
\nl && \qquad
- {5 \sqrt{3}}\eta\; \partial_\mu K^\dagger \partial_\nu\pi 
\partial_\rho\pi \partial_\sigma K + \dots \bigg]\epsilon^{\mu\nu\rho\sigma}
\nl
&\propto& \int\! d^4 x \, \partial_\mu \pi^+ \partial_\nu \pi^-
\partial_\rho \pi^0 K^\dagger \partial_\sigma K
\epsilon^{\mu\nu\rho\sigma} + \dots \,.  \end{eqnarray} 
Witten~\cite{Witten}
pointed out that this expression arises from a topological
construction in $D=5$.  It reflects the nontrivial homotopy group
$\pi_5(SU(3))=\bm{Z}$ and its coefficient is subject to quantization.
In QCD the WZ term locks the parity of the pion to that of spacetime, and it
describes allowed strong interactions such as $K \mybar{K} \rightarrow
\pi\pi\pi$, which are otherwise absent in a pure kinetic-term chiral
Lagrangian. 

In Little Higgs models the Higgs
field $H$ is identified with an object like the kaon, $H\sim K$. 
Upon electroweak
gauging, we thus expect to have a nontrivial WZW term, and by analogy
to the QCD chiral Lagrangian, new interactions can arise involving
gauge bosons and Little Higgs bosons.  In generic schemes we
might expect novel interactions like:
\begin{equation}\label{one} 
\int d^4x\, W_\mu^+ W_\nu^- Z_\rho^0 H_i^\dagger
\partial_\sigma H_j \epsilon^{\mu\nu\rho\sigma} \, + ... ,  
\end{equation} 
with multiple Little Higgs bosons $H_i$.

The discussion of such topological 
interactions requires the full gauging of the
Wess-Zumino-Witten term.  Witten originally initiated the program of
brute force gauging of the WZ term, which was developed by a number of
subsequent authors~\cite{Witten,kay,man,tye}.  For the cases where it
is applicable, we will employ the most transparent of these, the form
of Kaymacalan, Rajeev and Schechter (KRS)~\cite{kay}. The full WZW
term of KRS can be seen to descend from the Chern-Simons term and the
Bardeen counterterm \cite{bardeen} in a compactified pure Yang-Mills
theory of flavor in $D=5$~\cite{hillch}.

To retain intuition based on the familiar chiral Lagrangian of QCD, we
first consider a theory based on $SU(3)_L\times SU(3)_R \times
U(1)/SU(3)\times U(1)$, gauging an $SU(2)$ subgroup of $SU(3)_L$.
This is a simplified version of technicolor. The isovector techni-pions 
are eaten
to become longitudinal $W$ and $Z$ bosons, and physical techni-kaons
($\sim H$) and a techni-eta remain in the spectrum.  The resulting WZW
term indeed yields techni-kaon interactions of the form 
$ \epsilon^{\mu\nu\rho\sigma} W_\mu^+
W_\nu^- Z_\rho^0 H^\dagger \partial_\sigma H$.

We then turn to the simplest Little Higgs theories in which the
techni-pions of the previous example do not occur.  This ensures that
the $W$ and $Z$ bosons remain massless prior to electroweak symmetry
breaking.  These are generic Kaplan-Schmaltz (KS)
models~\cite{Kaplan:2003uc}.  They can be obtained by two physically
distinct approaches.  First, we consider reducing the $SU(3)_L\times
SU(3)_R \times U(1) / SU(3)\times U(1)$ QCD-like scheme to an
$SU(3)\times U(1)/SU(2)\times U(1)$ scheme by decoupling the
isovector techni-pions.  
In particular, we can ``eat and decouple'' the pions by
introducing an $SU(2)$ $A_R$ gauge field in a strong coupling limit. This
converts the usual unitary matrix chiral field into a {\em nonlinear
realization} of $SU(3)$ containing only the isodoublet Higgs (kaon)
and a singlet (eta).  This construction enforces the correct gauge 
transformation for a nonlinear realization, and dictates the proper form
of the covariant derivative.
Nonlinear realizations 
afford an interesting point of departure for the construction of Little 
Higgs models in general~\cite{2hill}. 
With the correct gauging established, we can then 
directly use the KRS form to obtain the full WZW
term for $SU(3)\times U(1)/SU(2)\times U(1)$ Little Higgs models.

Alternatively, we may contemplate directly the topological structure of the KS
models.  The pNGB's for $SU(3)\times U(1)/SU(2)\times U(1)$
are described by a complex triplet scalar field $\Phi$ with
$\Phi^\dagger\Phi = {\rm constant}$.  The fields thus live in a space
that is topologically equivalent to the five-dimensional unit sphere,
denoted $S^5$.  The topological interactions reflect the obvious but
nontrivial homotopy group $\pi_5(S^5) = \bm{Z}$, and are described by
the $SU(3)\times U(1)$-invariant form:~
\footnote{
Note that each term in eq.(\ref{onee}) involves five
``coordinates'' that locally parameterize an $S^5$. 
The term $\sim \eta dK^+ dK^-dK^0d\bar{K}^0$
corresponds to the submanifold of $SU(3)$ that is 
identified with $\Phi$.  
} 
\begin{equation}
\omega_{ABCDE} = -{i\over 8} \Phi^\dagger \partial_{[A } \Phi \partial_B\Phi^\dagger\partial_C\Phi\partial_D\Phi^\dagger \partial_{E] }\Phi  \,,
\end{equation}
which corresponds to the surface area of $S^5$ parameterized by
NGB's.  Under a local gauge transformation, $\delta \omega$ is a
total derivative in $D=5$ and can thus be gauged in $D=4$ to yield the
WZW term.  We describe the intricate procedure of gauging this
structure, and discuss the interesting question of equivalence of the
two approaches to the gauged topological action for $SU(3)\times
U(1)/SU(2)\times U(1)$.
We also touch on the related question of UV completions, noting that
the latter model can be viewed as arising from an underlying fermion
theory in which a triplet of techni-quarks $\psi_L$ condenses with a
singlet $q_R$, while the previous model arises from a condensate of two triplets 
of techni-quarks $\psi_L$ and $\psi_R$ with the pions eaten and decoupled.

While we presently sketch how all of this works, the full details will
be presented elsewhere~\cite{2hill}.  The above constructions provide
a building block for many composite Higgs scenarios.  Gauging
$SU(2)\times U(1)$ for a single complex triplet KS $\Phi$ field yields
a simple and intriguing model of a composite Higgs boson.
Gauging $SU(2)\times U(1)$ for two (or more) $\Phi_i$ fields yields a
multi-Higgs doublet model, while gauging the full $SU(3)\times U(1)$
for two $\Phi_i$ fields describes the Kaplan-Schmaltz Little Higgs
model.

When the symmetry breaking pattern respects an internal parity
operation the WZW term takes a special form that is identical to that
obtained in the QCD chiral Lagrangian.  This occurs, for example, in
Little Higgs models with ``T parity''~\cite{Cheng:2003ju}.  As an
application, we describe the main results for an $SU(5)/SO(5)$ Little
Higgs model~\cite{Arkani-Hamed:2002qy}.  We point out that the WZW
term is odd under the internal parity, and describes interactions
between a single ``T odd'' particle and standard model particles.

The WZW term contains interactions that are quantized, subject to
Adler-Bardeen nonrenormalization~\cite{AdlerBardeen}.  They are
suppressed by factors of $1/F$, with $F\sim 1$ TeV for typical Little
Higgs models, and occur with loop-order coefficients commensurate with
the underlying anomalies. Thus, they may be hard to detect.
Nonetheless, these interactions can provide a powerful discriminant of
underlying short-distance physics, and it is worth understanding what
effects can occur and determining whether they are suited to discovery
at the next generation of colliders, the LHC and the ILC.  Much of
this phenomenology is beyond the scope of the present paper, but will be 
developed elsewhere
\cite{2hill}.

\section{Illustration Based on Technicolor} 

To illustrate the procedures used in this analysis, we first consider
a chiral Lagrangian based on a QCD-like strong gauge group $SU(N_c)$,
containing $SU(3)$ flavor triplets of techni-quarks, $(\Psi_L,\Psi_R)$,
transforming in the fundamental representation with $N_c$ colors.  The
strong interaction results in a condensate $\langle \psi_L^i
\bar{\psi}_R^j \rangle \sim F^3 \delta^{ij}$, leading to an
$SU(3)_L\times SU(3)_R\times U(1) / SU(3)\times U(1)$ chiral
Lagrangian described by the $3\times 3$ unitary matrix field $U^{ij}
\sim \psi_L^i \bar{\psi}_R^j$.  We parameterize the field as $U =
\exp(2i {\tilde{\pi}}/F)$, where $\tilde{\pi}=\sum_{a=1}^8
\pi^a\lambda^a/2$ are Nambu-Goldstone bosons, transforming bilinearly
under $SU(3)_L\times SU(3)_R\times U(1)$ as:
\begin{equation}\label{Utransform}
U \rightarrow e^{i\epsilon_L} U e^{-i\epsilon_R} \,. 
\end{equation}
We presently turn off the standard model $U(1)_Y$
coupling constant, although it is straightforward to include it.  We
thus gauge an $SU(2)$ subgroup of $SU(3)_L$, with covariant
derivative:
\begin{equation}\label{Afield}
D_\mu U = \partial_\mu U - iA_\mu U \,, \qquad \qquad
A_\mu=\left( \begin{array}{cc} g W_\mu^a \frac{\sigma^a}{2} & 0 \\ 0 &
 0 \end{array}\right) \,. 
\end{equation}
In general such ``left-side'' gauging is anomalous, but
pure $SU(2)$ gauging is always anomaly free (we either ignore the
discrete Witten anomaly presently, or choose $N_c$ even).

The anomaly physics is contained in the gauged WZW term:
\begin{equation}\label{WZW}
\Gamma_{WZW} = \Gamma(U,A)  \,. 
\end{equation}
This can be obtained directly from the general form of $\Gamma(U,
A_L,A_R)$ in eq.(4.18) of KRS \cite{kay} (see eq.(\ref{WZWfull})
below), by setting $A_L = A$ and $A_R = 0$, where we use form
notation, \eg, $A = A_\mu dx^\mu$, $ABCD=
\epsilon^{\mu\nu\rho\sigma}A_\mu B_\nu C_\rho D_\sigma d^4x$, and
$dA=\frac12(\partial_\mu A_\nu - \partial_\nu A_\mu) dx^\mu dx^\nu$.  
A chiral current is defined as
$\alpha = dU U^\dagger = (2i/F) d\tp + ...$ \,.  Explicitly, we have:
\begin{eqnarray}
\label{final1}
\Gamma(U,A) & = & \Gamma_{0}(U) + 
\frac{  N_c}{48\pi^2} \int_{M^4} \Tr\left[ 
 {A}\alpha^3  -\frac{i}{2}({A}\alpha)^2
 +i(dA A + A dA)\alpha + {A}^3\alpha \right] \,,
\end{eqnarray}
where $M^4$ denotes four-dimensional spacetime ($M^5$ below denotes a
five-dimensional manifold with spacetime as its boundary).  We recall
that $\Gamma_0(U)$ is defined as the surface term of a globally
chirally invariant operator in $D=5$:
\begin{equation}\label{Gamma0}
{\Gamma}_{0}  = -{i N_c\over 240\pi^2} \int_{M^5} \Tr\left( \alpha^5\right)
= -\frac{i N_c}{240\pi^2}\int d^5x \,  \epsilon^{ABCDE}\;
\Tr \left( \alpha_A\alpha_B\alpha_C\alpha_D\alpha_E \right) \,. 
\end{equation}
$\Gamma_0$ is not manifestly local in four dimensions, but since
$d\Tr\alpha^5= \Tr\alpha^6=0$, it can be written as an expansion in
mesons in $D=4$.  Under a left-handed chiral transformation, $\delta U
= i\epsilon U$, we have $\delta\alpha = id\epsilon +i[\epsilon,
\alpha]$.  Using $\alpha^4=d\alpha^3$, it follows that $\Gamma_0$
shifts by an exact differential, a $D=5$ surface term,
\begin{equation}\label{brute}
\delta{\Gamma}_{0}  =  
\frac{ N_c}{48\pi^2}\int_{M^5}  
d \Tr \left( \epsilon \alpha^4 \right)
= 
 - \frac{N_c}{48\pi^2}\int_{M^4}  
 \Tr \left( d \epsilon\; \alpha^3 \right) \,. 
\end{equation}
This shift is compensated by the $D=4$ term in $\Gamma(U,A)$ involving
one gauge field.  The residual shift is cancelled by the term with two
gauge fields, and so on, leading
to eq.(\ref{final1}).~%
\footnote{ Alternatively, eq.(\ref{final1}) and eq.(\ref{WZWfull}) can
  be derived from the Yang-Mills Chern-Simons term and the Bardeen
  countertyerm in a $D=5$ Yang-Mills theory compactified on $S^1$
  where $A_5$ is identified with the mesons \cite{hillch}.  }

An important comment is in order concerning gauge invariance.  The
full WZW term of eq.(\ref{final1}) generates a vanishing anomaly in
the gauged $SU(2)$ subgroup, but nonvanishing anomalies in the global
currents, $a=4,...,8$.  That is, under $\delta U = i\epsilon U$ and
$\delta A = d\epsilon + i[\epsilon,A]$ we have $\delta \Gamma \propto
\Tr[\epsilon(dA\, dA - id A^3/2)]$.  This is the left-right symmetric
(or ``consistent'') form of the anomaly and it is not gauge covariant.
 It is always possible to add Bardeen's counterterm~\cite{bardeen} 
to bring the anomaly into the ``covariant'' form,
$\delta\Gamma \propto 3\Tr[\epsilon (dA - iA^2)^2]$.  However, the
coupling to NGB's is unaffected by the Bardeen counterterm, which is a
function only of gauge fields.
Expanding eq.(\ref{final1}) to leading order in NGB's, we see
that the NGB's participate in an interaction $\propto\Tr[\tp ( dA\,dA - id
A^3/2)]+...$~ which takes the form of the
consistent anomaly, and is superficially non-gauge invariant.  
 How, then, can we see that this gives
a bona-fide gauge invariant interaction?  The resolution is that, with
``left-side'' gauging of $U$, the gauge fields {\em always acquire a
  mass}.  Since the $SU(2)$ gauge anomalies vanish, we are free to
transform to unitary gauge to remove eaten NGB's.  The gauge fields
can then be expressed in the ``Stueckelberg'' form, $A^\prime =
V^\dagger ( A + id ) V$, where $V$ is the transformation to unitary
gauge that removes the NGB's from $U$.  The Stueckelberg fields
are gauge invariant. The residual physical
techni-mesons, expressed in the same gauge, couple to $A^\prime$, and
the interaction takes the form $\Tr[ \tp^\prime ( dA^\prime dA^\prime
- id A^{\prime 3}/2)] + ...$, which is manifestly gauge invariant. In
summary, ``left-side'' gauging produces NGB's coupled to the
consistent anomaly which is
a gauge invariant functional of covariant Stueckelberg fields.~%
\footnote{ For illustration, consider the hypothetical situation in
  which the usual Higgs mechanism does not operate, but where the
  left-handed quarks of some number of Standard Model generations, and
  hence the QCD chiral Lagrangian, is gauged with the usual $SU(2)$
  weak interactions.  As is well known, the $W$ and $Z$ bosons then
  eat a linear combination of the pions and acquire mass.  After
  transforming to unitary gauge, the remaining pseudoscalar mesons of
  QCD couple to the $W$ boson through the consistent, not the
  covariant, anomaly.  }

Let us examine some typical physical processes contained in
eq.(\ref{final1}).  A convenient choice of $SU(3)$ coordinates around
$U=1$ is:
\begin{equation}
\label{Umatrix}
U = \left( 
\begin{array}{cc}
e^{i\eta/F} \sqrt{1 - H H^\dagger/F^2} \,\, \hat{U} & H/F \\
-e^{-i\eta/F} H^\dagger \hat{U}/F & e^{-2i\eta/F}\sqrt{1-H^\dagger H/F^2}
\end{array} 
\right) \,,
\end{equation}
where $\hat{U} = \exp(2i\hat{\pi})$ is an $SU(2)$ matrix containing
the techni-pions, $\hat{\pi} = \sum_{a=1}^3 \hat{\pi}^a\sigma^a/2$,
$\eta$ is a real phase and $H$ is a complex iso-doublet (techni-kaon).  
Here $H
H^\dagger $ denotes a dyadic product.  Under an $SU(2)_L$
transformation, $\epsilon_L = {\rm diag}(\hat{\epsilon}, 0)$ and
$\epsilon_R=0$ in eq.(\ref{Utransform}), we have $H\to
e^{i\hat{\epsilon}} H$, $\hat{U}\to e^{i\hat{\epsilon}} \hat{U}$ and
$\eta \to \eta$.
 
We presently focus attention on terms involving $H$ through second
order in $1/F$.  Thus the chiral current becomes:
\begin{equation}\label{alpha}
\alpha =  \left( \begin{array}{cc} \frac{-1}{2F^2} H
\stackrel{\leftrightarrow}{d} H^\dagger 
& \frac{1}{F} dH
\\  \frac{-1}{F} dH^\dagger 
 &   
 \frac{-1}{2F^2} H^\dagger \stackrel{\leftrightarrow}{d} H \end{array} \right) \,, 
\end{equation}
and we find the WZW interactions of eq.(\ref{WZW}) take the form:
\begin{eqnarray}
\label{result1}
\Gamma_{WZW} & = &  - { i g^2N_c\over 192 \pi^2 F^2 } \int d^4 x\; 
\epsilon^{\mu\nu\rho\sigma} 
 \left( 
Z_\mu^0 \partial_\nu Z_\rho^0 + W_\mu^-\partial_\nu W_\rho^+ +
W_\mu^+\partial_\nu W_\rho^- 
-  {3 i\over 2} g Z_\mu^0 W_\nu^+ W_\rho^- \right ) \nl
&& \qquad\qquad \times  \left(H^0 \stackrel{\leftrightarrow}{\partial}_\sigma H^{0*} 
+ H^+ \stackrel{\leftrightarrow}{\partial}_\sigma H^-\right)  + \dots \,.
\end{eqnarray}
The kinetic term allowed by $SU(3)_L\times SU(3)_R \times U(1)$
invariance is $F^2 {\rm Tr}|D_\mu U|^2$.  In this ``technicolor''
scheme, the $W$ and $Z$ bosons eat the $\hat{\pi}$ degrees of freedom,
acquiring a common mass $gF$.  We can transform to unitary gauge,
$\hat{U}\rightarrow 1$, leaving physical fields $\eta$ and $H$, and
eq.(\ref{result1}) describes the anomalous interactions of these
physical fields with the massive Stueckelberg $W$ and $Z$ gauge
bosons.  The WZW term for this technicolor scheme contains the
interesting physical processes $e^+e^- \rightarrow Z^*\rightarrow
W^+W^- (H^0 H^{0*} + H^+H^-)$ and $e^+e^- \rightarrow Z^*\rightarrow
Z^0 (H^0 H^{0*} + H^+H^-)$, with amplitudes that count the number of
underlying techniquark colors, $N_c$.

Eq.(\ref{result1}) does not describe a Little Higgs theory,
but we can deform this technicolor theory
to imitate a Little Higgs scheme by restricting the
kinetic term to the form $F^2 \Tr|D_\mu U P' |^2$
with the projection matrix $P' = {\rm diag}(0,0,1)$.  Doing so 
blocks the $W$ and $Z$ from acquiring mass by
projecting out their longitudinal
coupling with the isovector techni-pions, but unfortunately 
leaves the $\hat{\pi}$ as nonpropagating auxiliary fields in the theory.
We will see subsequently that removing the unphysical techni-pions 
enforces that $U$ transform as a {\em nonlinear realization} of $SU(3)$.  
The problem for Little Higgs theories is
therefore to construct the WZW term either by attacking directly and
rederiving the full WZW action in terms of a restricted manifold of
NGB's, or by adapting the above familiar form of the WZW term to the
case of $U$ treated as a nonlinear realization.  We turn to this issue
in the next sections, and derive the topological physics of bona-fide
Little Higgs models.

\section{WZW term for models involving $SU(3)/SU(2)$}

A set of ``simple'' Little Higgs models are based on $SU(3)\times U(1)
/SU(2)\times U(1)$, (or more generally $SU(n)\times U(1)
/SU(n-1)\times U(1)$).  These were introduced by Kaplan and Schmaltz
(KS) \cite{Kaplan:2003uc} .  We begin by considering ``one half'' of
such an $SU(3)/SU(2)$ Little Higgs model, described by a single scalar
field $\Phi$ which transforms as a triplet under $SU(3)$.

We can view the KS model as arising from a UV completion scheme in
which $\Phi$ is a bound state $\phi^i \sim \psi^i_{L} \bar{q}_{R} $,
where $\Psi_L$ is a flavor triplet and $q_R$ a singlet.  The fermions
transform in the fundamental representation of a color group $SU(N_c)$
and we assume that a $(3,1)$ condensate forms with $ \VEV{\Psi_L \bar{q}_R}
\sim (0,0,F^3)^T$.  We have constructed such UV completions and will
discuss their full content elsewhere.  The unbroken $SU(2)\times U(1)$
subgroup is anomaly free, with a vector-like $U(1)$ current $\propto
\overline{\psi}_L^3\gamma_\mu\psi_L^3 + \bar{q}_R\gamma_\mu q_R $.  This
subgroup can thus be identified with the electroweak gauge group of
the standard model. We can obtain the topological interactions
by deriving the WZW term directly as a functional of $\Phi$, which amounts
to gauging the sphere $S^5$ defined by 
$\Phi^\dagger \Phi = 1$.

Alternatively, the KS model can be viewed as arising from
a UV completion with an $SU(3)_L\times SU(3)_R$ chiral structure, a
$(3,\bar{3})$ condensate of the form $U\sim \psi_L\bar{\psi}_R$. 
The iso-vector pions are removed by gauging on the right with
a strongly coupled iso-vector gauge field $A_R$. $A_R$ eats the pions,
and becomes a functional of $H$, $\eta$ and the remaining gauge fields $A_L$,
while $U$ becomes a nonlinear realization with a modified
covariant derivative. We begin with the nonlinear realization.

\subsection{ Nonlinear Realization} 

For the low energy effective Kaplan-Schmalz 
theory we can write $\Phi = U \times \langle \Phi \rangle$,
where $\langle \Phi \rangle$ is a constant vector, and 
$U$ is a unitary matrix function of the five (or $2n+1$) NGB's.  
A convenient choice of coordinates for $\langle \Phi \rangle$ and $U$ is
\begin{equation}
\label{higgs}
\langle \Phi \rangle = \left(\begin{array}{c} 0 \\ 0 \\ 1 \end{array} \right) \,,
\qquad \qquad
U = 
\exp\left[ {i\over F}\left(\begin{array}{cc} \eta I_2/\sqrt{3} & H  
\\ H^\dagger & -2\eta/\sqrt{3} 
\end{array}\right)\right] \,.
\end{equation}
$H$ has the usual standard model quantum numbers of the Higgs, and
$\eta$ is a standard model singlet.  For simplicity, we will ignore
the $U(1)$ factor in the remainder of this section, although it is
straightforward to include it.

The critical element of this formalism is that $U$ transforms as a
{\em nonlinear realization} of $SU(3)$~\cite{Coleman:1969sm}.  That
is,
\begin{equation}\label{nonlinear} 
U \rightarrow e^{i\epsilon}U e^{-i\epsilon'} \,,
\end{equation}
where $\epsilon \in SU(3)$ and $\epsilon'$ is a matrix function of
$\epsilon$ and $U$, belonging to the unbroken $SU(2)$ subgroup, the
upper left block with our choice of coordinates.  Hence, while $U$
contains only the five degrees of freedom described by $H$ and $\eta$,
we can implement the full $SU(3)$ transformation, albeit nonlinearly.

The key challenge with nonlinear realizations is that the ordinary
derivative, $dU$, is not covariant under global transformations with
constant $\epsilon$, owing to the spacetime dependence of
$\epsilon'(\epsilon, U(x))$.  We therefore need to construct a
covariant derivative that involves the gauge field $A_L$ and acts on
$U$ so that $DU \rightarrow e^{i\epsilon}DU e^{-i\epsilon'}$.  Such
a covariant derivative can be written as follows:
\begin{equation}
\label{cov1}
DU = d U - i A_L U + i  U A_{R} \,, 
\qquad A_{R} = P U^\dagger(A_L + i d )U P
- {1\over n-1}{\rm Tr}\left[ P U^\dagger(A_L + i d )U P \right] \,. 
\end{equation}
$A_R$ is the projection of $U^\dagger(A_L+id)U$ onto the unbroken
subgroup.  Here we have defined the projection matrix 
\begin{equation}
P = \left(\begin{array}{ccc} 1 \\ & 1 \\ && 0 \end{array} \right) \,.
\end{equation} 
Under the gauge transformation eq.(\ref{nonlinear}), with
\begin{equation}\label{trans0}
\delta A_L = d\epsilon + i[\epsilon,A_L],
\end{equation}  
$A_R$ transforms as:
\begin{equation}
\label{trans1}
\delta A_R = d\epsilon' +i[\epsilon', A_R] \,,
\end{equation}
ensuring that $DU$ is covariant.

The nonlinear realization eq.(\ref{nonlinear}) embeds the NGB's from
$SU(3) \to SU(2)$ inside a larger manifold corresponding to
$SU(3)_L\times SU(3)_R\to SU(3)$.  We can describe the removal of the
extra pionic degrees of freedom in physical terms as follows.  We begin by
treating $A_L$ and $A_R$ as independent gauge fields, where $A_L$ is a
general $SU(3)$ matrix field, and $A_R$ belongs to the unbroken
$SU(2)$ subgroup.  Correspondingly, the $\epsilon$ and $\epsilon'$ in
the transformation eq.(\ref{nonlinear}) are independent rotations.  Now
suppose that the $A_R$ kinetic term vanishes, corresponding to very
strong coupling, \ie, $ (-1/g_R^2)\Tr F_{R\mu\nu} F_{R}^{\mu\nu}
\rightarrow 0$.  $A_R$ then becomes an auxiliary field with equation
of motion determined by the $SU(3)_L\times SU(3)_R$ chiral-invariant
kinetic term, $F^2 \Tr|D_\mu U|^2$.  Noting that $P A_R P = A_R$ and
${\rm Tr}(A_R) =0$, we obtain precisely the locking condition
eq.(\ref{cov1}) as the solution for $A_R$ as a function of $A_L$ and
$U$.  This allows us to ``eat and decouple'' the unwanted 
isovector NGB's
in $U$.  Using the gauging of
eq.(\ref{cov1}) and expanding the resulting kinetic term in powers of $1/F$ we
see that $(F^2/2) |D_\mu U|^2 \to |D_\mu H|^2 + (\partial_\mu \eta)^2
+ \dots$~, and we are therefore dealing with a Little Higgs theory. Thus,
in a sense, a Little Higgs theory is just a technicolor theory with 
the usual chiral field $U$ replaced by a nonlinear realization.

For any chiral theory based on a unitary matrix $U$ transfoming as
$U\rightarrow e^{i\epsilon} U e^{-i\epsilon'}$ and gauge fields $A_L$
and $A_R$ that likewise transform under $\epsilon$ and $\epsilon'$ as
in eqs.(\ref{trans0}) and eq.(\ref{trans1}), the gauged WZW term is given
explicitly by KRS eq.(4.18):
\begin{align}
\label{WZWfull}
&{\Gamma}_{WZW}(U, A_L, A_R)  = \Gamma_0(U) \;  + \frac{N}{48\pi^2}\Tr\int_{M^4} \bigl\{  
({A}_L\alpha^3 + {A}_R\beta^3 ) 
-\frac{i}{2}[({A}_L\alpha)^2 - ({A}_R\beta)^2 ] 
\nonumber \\
&\quad
+i \big[ (dA_LA_L + A_L dA_L)\alpha+ (d{A}_R{A}_R +{A}_RdA_R)\beta \big]
+({A}^3_L \alpha+{A}^3_R \beta)
\nonumber \\
&\quad
+i(A_L U {A}_R U^\dagger \alpha^2 - {A}_R U^\dagger A_L U \beta^2)  
+i ( dA_R dU^\dagger A_L U - dA_L dU A_R U^\dagger  )
\big] 
\nonumber \\
&\quad
- (dA_L A_L + A_L dA_L )U A_R U^\dagger +(dA_R A_R + A_R d A_R) U^\dagger A_L U
\nonumber \\
&\quad
-i(A_L U  A_R U^\dagger A_L \alpha + A_R U^\dagger A_L U A_R \beta  ) 
+i\big[ A_L^3 U  A_R U^\dagger - A_R^3 U^\dagger A_LU - 
\half (UA_RU^\dagger A_L)^2 
\bigr\} \,, 
\end{align}
where $N$ is an integer; \eg, in the QCD chiral Lagrangian, $N=N_c=3$
is the number of colors.  Here $\alpha = dU U^\dagger$ and $\beta =
U^\dagger dU$.  The function $\Gamma_0(U)$ is given by eq.(\ref{Gamma0}),
which in four dimensions reads
\begin{equation}
{\Gamma}_{0}(U) = 
{2N\over 15\pi^2 F^5} \int_{M^4} {\rm Tr}\left[ \tp (d\tp)^4 \right] 
+ ...  \,. 
\end{equation}

In the present case we need only substitute the representation of $U$
given in eq.(\ref{higgs}) and the locking of $A_R$ to $A_L$ and $U$
given in eq.(\ref{cov1}).  We can then expand to a given order in
$1/F$ to obtain the topological interactions of the mesons and gauge
fields.  We are presently ignoring $U(1)$ factors, and identify the
unbroken $SU(2)$ subgroup with electroweak gauge interactions
(we'll also presently ignore the $U(1)_Y$ gauge subgroup):
\begin{equation}
A_L = \left(\begin{array}{cc} W & 0 \\ 0 & 0 \end{array} \right) \,, 
\end{equation}
where $W = gW^a\sigma^a/2$.  Defining
\begin{equation}
A_R = A_L +   \left(\begin{array}{cc}  
\hat{A}_R  &
0  \\  0 &  0 
\end{array}\right) \,, 
\end{equation} 
we find to second order in $1/F$:
\begin{eqnarray}\label{U} 
\hat{A}_R = - \frac{1}{2F^2} \left( \{ W, H H^\dagger \} - H^\dagger W H \right) 
+\frac{i}{2F^2} \left( H \stackrel{\leftrightarrow}{d} H^\dagger
+ \frac12 H^\dagger \stackrel{\leftrightarrow}{d} H \right) \,, 
\end{eqnarray}
and
\begin{eqnarray}
\label{alph}
\alpha &=& \left(\begin{array}{cc}  
{i\over \sqrt{3}F}d\eta -\frac{1}{2F^2} 
H \stackrel{\leftrightarrow}{d} H^\dagger \quad 
& \quad {i\over F}\left( 1+{\sqrt{3}i\over {2F}}\eta\right) dH + {\sqrt{3}\over
{2F^2}} H d\eta 
\\ 
{i\over F}\left( 1- {\sqrt{3}i\over {2F}}\eta\right) dH^\dagger 
- {\sqrt{3}\over {2F^2}} H^\dagger d\eta 
 & 
-{2i\over \sqrt{3F}} d\eta 
 -\frac{1}{2F^2} H^\dagger  \stackrel{\leftrightarrow}{d} H 
\end{array}\right) \,. 
\end{eqnarray}
Here $A\stackrel{\leftrightarrow}{d}B \equiv
A\stackrel{\rightarrow}{d}B - A \stackrel{\leftarrow}{d}B$.  The
expansion for $\beta$ is given (to all orders in $1/F$) by
substituting $H\to -H$ and $\eta\to -\eta$ into $-\alpha$.

The leading WZW interactions involving $W$ and NGB's appear at order
$1/F$:
\begin{equation}\label{nonlinearWZW}
\Gamma_{WZW} = 
{-N \over 8\pi^2\sqrt{3}F } 
\int_{M^4} {\eta} {\rm Tr}( F_W^2 ) + \dots   \,, 
\end{equation}
where $F_W = dW-iW^2$.  As the result of $SU(2)$ matrix identities, no
additional interactions appear through order $1/F^2$ from the action in
eq.(\ref{WZWfull}) involving just $W$ and $H$. In more general gauging
with $U(1)_Y$, and generalizations to multiple chiral fields
$\Phi_i \sim U_i P^\prime$ there can occur in principle at this order 
other interesting operators such as
$H^\dagger F_W H F_Y$ which lead
to processes such as $e^+e^- \rightarrow (\gamma^*,Z^*) \rightarrow (h^0, A^0) 
+(Z,\gamma,W^+W^-)$ ($A^0$ is a CP-odd Higgs boson).
There also appear operators 
at this order that are separately
gauge-invariant and chirally invariant:
\begin{equation}\label{GI}
\Gamma_{GI} = \int_{M^4} i r {\rm Tr}(F_L U F_R U^\dagger) 
= \int_{M^4} - {2 i r\over F^2} H^\dagger F_W^2 H + \dots 
\,,  
\end{equation}
with a coefficient $r$ that is not quantized, and is sensitive to the
details of the underlying UV completion theory. Multi-Higgs topological
interactions of $W$ and $H$ occur at order $1/F^4$:
\begin{equation}\label{resultx}
\Gamma_{WZW} \supset
{N\over 96\pi^2F^4} \int_{M^4} 
 H^\dagger (DH) \, H^\dagger F_W (DH) + (DH^\dagger)H\, (DH^\dagger) F_W H \,. 
\end{equation} 
This operator can in principle lead to processes such as:
 $e^+e^- \rightarrow (\gamma^*,Z^*) \rightarrow h^0 + (Z,\gamma) $
and will generalize in other schemes \cite{2hill}.

Additional quantized, topological, interactions will occur when the
mesons couple to gauge fields for broken symmetry generators.  This
happens in Little Higgs models, where anomaly cancellation occurs
between different sectors that are connected only by weak gauge
interactions.  Before discussing this phenomenon in more detail, we
describe in the next section an alternative, and more direct approach
to the WZW term for $SU(3)/SU(2)$.

Finally, let us remark that the construction here based on
$SU(3)_L\times SU(3)_R/SU(3)$ relied on being able to gauge the
$SU(2)$ subgroup of $SU(3)_R$ and remove the unwanted pion degrees of
freedom.~%
\footnote{ The global structure of the NGB manifold in this case (a
  submanifold of $SU(3)$) allows an arbitrary integer coefficient,
  $N$, for the WZW term.  However, the natural UV theory in this
  physical picture has $A_R$ coupled to a right-handed fermion
  doublet.  For this theory to be free of discrete
  anomalies~\cite{Witten:1982fp} there should be an even number of
  such doublets, and therefore an even number of colors.  This is
  related to a factor of 2 advocated by the authors of
  \cite{D'Hoker:1994ti} in cases where the unbroken subgroup $H$, \eg\,
  $H=SU(2)$, has $\pi_4(H)=\bm{Z}_2$.  With the global structure
  defined by the direct construction based on $S^5$, a factor of
  $2$ is enforced automatically.  }  
In the more general case of
$SU(n)/SU(n-1)$ based on $SU(n)\times SU(n)/SU(n)$, the unbroken
subgroup is anomalous,
and so cannot be gauged without adding more structure.~%
\footnote{ More precisely, a gauged WZW term can be written down for these
  cases, but the variation of the action results in an anomaly
  expression containing mesons, so that the theory does not have a
  simple UV completion in terms of constituent fermions.  }  
For example, if the anomaly of colored fermions is cancelled by massless
right-handed leptons,
\begin{equation}
\Delta {\cal L}_{\rm lepton} =  \bar{\ell }_R (i\slash{\partial} 
+ \slash{A}_R ) \ell_R \,,
\end{equation}
then $A_R$ in eq.(\ref{cov1}) is modified to
\begin{equation}
A_{R\mu}^a  \to A_{R\mu}^a - {1\over F^2} \bar{\ell}_R t^a \gamma_\mu \ell_R \,.
\end{equation}
Substituting this new locking condition for $A_R$ as a function of
$A_L$ and $U$ into eq.(\ref{WZWfull}) and setting $A_L=0$ yields an
ungauged action involving the leptons and the remaining mesons.  The
corresponding gauged action can then be constructed by ``brute-force''
gauging, as described after eq.(\ref{brute}).

\subsection{Gauging the Five Sphere}

A more direct method for obtaining the WZW term for KS models is to
notice that when the meson fields are described by a vector $\Phi =(
\phi^1,\phi^2,\phi^3)^T$ with
\begin{equation}
\Phi^\dagger \Phi = (\Re\phi^1)^2 + (\Im\phi^1)^2 + (\Re\phi^2)^2 + (\Im\phi^2)^2 +
(\Re\phi^3)^2 + (\Im\phi^3)^2  = 1 \,,
\end{equation}
the field lives on a manifold with very simple global topological
structure, the five-dimensional sphere, $S^5$.  The relevant
topological facts are that $\pi_4(S^5)=0$, guaranteeing that the
construction of a WZW term is possible, and $\pi_5(S^5)=\bm{Z}$,
guaranteeing that the result is nontrivial.  There is a unique
five-form on the five-sphere that is invariant under global
$SU(3)\times U(1)$ rotations, namely the volume element of the sphere,
\begin{equation}\label{omega}
\omega = -{i\over 8}\Phi^\dagger d\Phi d\Phi^\dagger d\Phi d\Phi^\dagger d\Phi \,.
\end{equation}
In analogy with the construction of $\Gamma_0(U)$ in eq.(\ref{Gamma0}),
we consider the topological action
\begin{equation}\label{sphereWZ}
\Gamma_0 = {N\over \pi^2} \int_{M^5} \omega \,. 
\end{equation}
Since any five-form on a five-dimensional manifold is closed,
$d\omega=0$, this action can be written as an expansion in Goldstone
bosons in $D=4$.  Using that the area of the five-sphere is $\pi^3$,
the coefficient satisfies the quantization condition displayed in
eq.(\ref{sphereWZ}), with $N$ an {\it even} integer.

Gauging the topological action is more tedious than in the familiar
$SU(N)\times SU(N)/SU(N)$ case.  At a practical level, the gauging
begins by noticing that the variation of the ungauged action yields
\begin{equation}\label{spherevar}
\delta\Gamma_0 = {N\over 8\pi^2}\int_{M^4} 
\left( \Phi^\dagger d\epsilon d\Phi + d\Phi^\dagger d\epsilon \Phi -d\epsilon_0
\Phi^\dagger d\Phi  
\right) d\Phi^\dagger 
d\Phi \,,
\end{equation}
where $\epsilon$ and $\epsilon_0$ are the $SU(3)$ and $U(1)$
components in the variation of $\Phi$, respectively:
\begin{equation}
\Phi\to e^{i (\epsilon+ \epsilon_0) } \Phi \,. 
\end{equation}
Eq.(\ref{spherevar}) has made use of the fact that for fields confined
to the
five-sphere,~%
\footnote{ For a geometrical description of such identities, see
  \cite{Hull:1990ms}.  }
\begin{equation}
\begin{array}{c}
d(\Phi^\dagger \lambda^a \Phi ) (d\Phi^\dagger d\Phi)^2 = 0 \,, \\
\left[ \Phi^\dagger \lambda^a \Phi d\Phi^\dagger d\Phi 
-2 d(\Phi^\dagger \lambda^a \Phi )\Phi^\dagger d\Phi 
+2 d\Phi^\dagger \lambda^a d\Phi \right] d\Phi^\dagger d\Phi = 0 \,.  
\end{array}
\end{equation}
The variation eq.(\ref{spherevar}) is compensated by the term with one
gauge field,
\begin{equation}
\Gamma_1 = {N\over 8\pi^2} \int_{M^4} \left( A_0 \Phi^\dagger d\Phi - \Phi^\dagger A d\Phi - d\Phi^\dagger A \Phi \right) 
d\Phi^\dagger d\Phi \,. 
\end{equation}
The residual shift is cancelled by a term with two gauge fields, and
so on.  Explicit expressions for the full gauged action will be
presented elsewhere~\cite{2hill}.  Important aspects of the analysis
are the non-uniqueness of the gauged WZW term, and a restrictive
interpretation in terms of underlying fermions, which we turn to
presently.

The complete result for the gauged WZW term leads to an anomalous
gauge variation:
\begin{align}\label{eq:4dvar}
  \delta\Gamma_{WZW} &= -{N\over 24\pi^2} \int_{M^4} {\rm Tr}\left\{
  \left(\epsilon-\frac{\epsilon_0}{2} \right) 
\left[ \left(dA - \frac12 dA_0 \right)^2 - {i\over 2}d\left(A-\frac12 A_0\right)^3 \right] 
\right\}  + \frac{27}{8} \epsilon_0 (dA_0)^2 \,.
\end{align} 
Although the construction has made no mention of fermions, if
interpreted in terms of an underlying fermion theory, this variation
corresponds precisely to a triplet of left-handed fermions and a
single right-handed fermion, transforming under $\epsilon \in SU(3)$
and $\epsilon_0 \in U(1)$ as
\begin{equation}\label{fermi}
\Psi_L \to e^{i\epsilon - {i\over 2}\epsilon_0 } \Psi_L \,,\quad
q_R \to e^{-{3i\over 2} \epsilon_0 } q_R  \,.  
\end{equation}
Note that this is precisely the combination of $U(1)$ transformations
that is not broken by color anomalies.  The quantized coefficient
corresponds to an {\it even} number of colors $N=N_c$ in the color
group.

As mentioned above, the unbroken $SU(2)\times U(1)$ subgroup is
anomaly free, and can
be identified with the electroweak gauge group of the standard model.~%
\footnote{ Having identified an anomaly-free embedding of the unbroken
  subgroup, and the anomalous gauge variation of the full action, we
  can work in the opposite direction to reconstruct the gauged WZW
  term by integration~\cite{Chu:1996fr}.  The non-uniqueness of the
  WZW term will be reflected in a choice of boundary condition for
  this construction.  } The symmetry breaking corresponds to a nonzero
VEV $\langle \Psi_L \bar{q}_R \rangle \sim (0,0,F^3)^T$.  For
simplicity, we ignore the $U(1)$ factor in the remainder of this
section, and use the choice of coordinates of eq.(\ref{higgs}).  The
leading WZW interactions are identical to eq.(\ref{nonlinearWZW}).
There are again operators that are invariant under local $SU(3)\times
U(1)$ transformations.  Suppressing $U(1)_Y$, 
\begin{equation}\label{GIsphere}
\Gamma_{GI} = \int_{M^4} c\, \Phi^\dagger (dA - iA^2)^2 \Phi 
+ \dots  \,,   
\end{equation}
where $c$ is a number sensitive to the UV completion theory, and 
the ellipsis denotes additional terms 
such as $[\Phi^\dagger (dA-iA^2) \Phi]^2$ relevant at $\order(1/F^4)$.   
When viewed as deriving from an $SU(3)_L\times SU(3)_R/SU(3)$ theory,
parity arguments can be invoked to single out a unique choice of these
gauge-invariant operators~\cite{kay}, \eg\, $r=0$ in eq.(\ref{GI}), leading
to particular values of the coefficients in eq.(\ref{GIsphere}).
Sensitivity to the global structure of the field space of NGB's,
equivalently, sensitivity to the UV completion theory, is reflected in
the undetermined coefficients in eqs.(\ref{GI}) and (\ref{GIsphere}).

It is interesting to investigate further what novel features of
$SU(3)/SU(2)$ allow an underlying fermion theory to produce a
low-energy theory containing just scalar NGB's and gauge fields, and
in particular why a similar construction is not possible for general
$SU(n)/SU(n-1)$.  At the meson level, the nonlinear realization
construction required an anomaly-free gauging of $A_R$ to ``eat and
decouple'' the extra pions; this singled out $SU(2)$ as the unbroken
symmetry group (with an even number of colors in the strong color
group).  The direct construction required the form $\omega$ in
eq.(\ref{omega}) to be closed, $d\omega=0$, which is true for $S^5$, but
not for general $S^{2n-1}$ (notice that $\pi_5(S^5)=\bm{Z}$, but
$\pi_5(S^{2n-1})=0$ for $n>3$).  At the fermion level, we expect in
general that the condensate $\langle \Psi_L \bar{q}_R \rangle \sim
(0,0,\dots,0,F^3)^T$ will leave $n-1$ massless fermions $\psi_L^{i}$,
$i=1\dots (n-1)$, in the low-energy spectrum.  For $n>3$, these
fermions enable the low-energy theory to reproduce the full nonabelian
anomaly of the underlying UV theory.  For the special case of $n=3$,
such fermions are not mandated by anomaly matching.  For example, when
$N_c=2$, we observe that it is possible to write the operator
\begin{equation}\label{UV}
\epsilon_{ijk} \epsilon^{ab}\epsilon^{\alpha\beta}
 \psi^i_{a\alpha} \psi^j_{b\beta} \, \epsilon^{cd}\epsilon^{\gamma\delta} \psi^k_{c\gamma} {q}_{d\delta}  + h.c. \,, 
\end{equation}
where $i,j,\dots = 1..3$ are flavor indices, $a,b,\dots = 1..2$ are
color indices, and $\alpha,\beta,\dots=1..2$ are Lorentz indices in
the $(1/2,0)$ representation of the Lorentz group.  The operator
eq.(\ref{UV}) is invariant under Lorentz and color $SU(N_c=2)$
transformations, and also under the flavor $SU(3)\times U(1)$
transformation in eq.(\ref{fermi}).  When $\Psi_L \bar{q}_R $ develops a
VEV, the operator becomes a (Majorana) mass term for the remaining
$\psi_L^{i}$ fermions, removing these degrees of freedom from the
low-energy spectrum.

\section{Little Higgs Models}

Having constructed the full WZW term, we consider applications to
realistic Higgs models.  As mentioned in the introduction, the
$SU(3)/SU(2)$ symmetry breaking pattern can be applied to a number of
scenarios; we focus on the KS model here, concentrating attention on
predictions that are independent of the undetermined coefficients in
eqs.(\ref{GI}) and (\ref{GIsphere}).  We recall that this model in its 
simplest form
consists of two $\Phi_i$ fields, with aligned vacuum expectation values.
The gauge fields, suppressing $U(1)$ factors, take the form
\begin{equation}
A = \left( \begin{array}{cc} W & C \\ C^\dagger & 0 \end{array} \right) \,. 
\end{equation}
Concentrating on interactions involving $H$, (neglecting $\eta$), the
kinetic terms are
\begin{equation}
F^2 |D_\mu\Phi_i|^2 
= |D_\mu H_i|^2 - F C_\mu^\dagger D_\mu H_i - F (D_\mu H_i^\dagger) C_\mu + 
F^2 C_\mu^\dagger C_\mu + \dots 
\dots \,,
\end{equation}
where $i=1,2$ and $D_\mu H = (\partial_\mu - i W_\mu)H$.  Besides the
usual kinetic term for $H$, this expression contains an $F$-scale mass
for $C$.  The NGB's from the symmetry breaking at scale $F$ are eaten
by the $C$ bosons, and the physical Higgs fields appear as, $H_1 = H$,
$H_2 = -H$ (similarly, $\eta_1= \eta$, $\eta_2 = -\eta$).  Terms
containing an odd number of meson fields thus cancel in the sum,
\begin{equation}
{\cal L}_K = (F^2/2) |D_\mu\Phi_1|^2  + (F^2/2) |D_\mu \Phi_2|^2 
= |D_\mu H|^2 + F^2 C^\dagger_\mu C_\mu + \dots \,. 
\end{equation}
The same cancellation will occur with the subleading even-parity
``Gasser-Leutwyler'' operators~\cite{Gasser:1984gg} (operators not
containing an epsilon symbol $\epsilon^{\mu\nu\rho\sigma}$) if the
strong-interaction physics is the same in both sectors.  In order that
anomalies cancel between the $\Phi_1$ and $\Phi_2$ sectors 
(at the fermion level, 
$\Phi_1 \sim \Psi_L \bar{q}_R$, $\Phi_2 \sim \Psi_R \bar{q}_L$), 
the opposite cancellation must occur with odd parity operators such as the
WZW term (operators containing an epsilon symbol)---surviving
interactions involve an {\it odd} number of meson fields.  Thus for
example, the interaction eq.(\ref{nonlinearWZW}) involving $\eta$ will
survive (with a factor of $2$ from the sum of the two sectors).  The
leading terms involving $H$ also occur at order $1/F$, and can be
obtained either from the nonlinear realization approach, or from the
direct gauging of $S^5$:
\begin{equation}\label{result2}
\Gamma_{WZW} \supset {N \over 16\pi^2 F} \int_{M^4} \left[ (D H^\dagger) F_W C - C^\dagger F_W DH \right]\,.
\end{equation}
This expression is for one sector (a factor of two will appear in the
sum of the two sectors).  It is manifestly gauge-invariant under
electroweak $SU(2)$.  The $SU(3)$-invariant odd-parity
Gasser-Leutwyler operators in eqs.(\ref{GI}) or (\ref{GIsphere})
contribute only to the orthogonal combination, eq.(\ref{result2}) with
the relative minus sign replaced by a plus sign.  The new interaction
eq.(\ref{result2}) would contribute to the process
$e^+e^- \to Z^* \to h^0 C^0$ (note that this is the analog
in QCD of the process $e^+e^- \rightarrow \rho \rightarrow KK^*$).

\section{WZW term for models with an internal parity}

The Lagrangian eq.(\ref{WZWfull}) can be used to
describe general symmetry breaking patterns, via reduction to
nonlinearly realized symmetries acting on submanifolds of a larger 
space.  For example, we obtained the WZW term for $SU(3)/SU(2)$ 
by embedding the NGB's inside a full $SU(3)\times SU(3)/SU(3)$ multiplet.  
A further simplification occurs for general models in which the symmetry
breaking pattern respects an internal parity operation, and eq.(\ref{WZWfull})
applies also to these cases.  The $SU(N)_L\times SU(N)_R/SU(N)$ 
QCD chiral Lagrangian is one example.  Another example is the 
class of Little Higgs models containing an internal parity.  
We examine here the structure of the WZW term for this case.

We recall that by suitable choice of coordinates, the action of an
element of the full symmetry group can be defined to act, at least
locally, on the Nambu Goldstone bosons as~\cite{Coleman:1969sm}
\begin{equation}\label{eq:nonlinear}
e^{i\pi} \to e^{i\pi^\prime} = e^{i\epsilon} e^{i\pi} e^{-i\epsilon^\prime(\epsilon,\pi)} \,.
\end{equation}
Here $\pi = \sum_a \pi^a t^a_A$ parameterizes the spontaneously broken
``axial'' symmetry generators, and $\epsilon^\prime = \sum_a
\epsilon^{\prime a} t^a_V$ is the combination of unbroken ``vector''
symmetry generators which ensures that
$e^{i\epsilon}e^{i\pi}e^{-i\epsilon^\prime}$ can be expressed as
$e^{i\pi^\prime}$ for some $\pi^\prime$.  Now suppose that the
transformation:
\begin{equation}
t^a_V \to R(t^a_V) = t^a_V \,, \quad t^a_A \to R(t^a_A)= -t_A^a 
\end{equation}
preserves the group structure, i.e. $[t_V, t_V] \sim t_V$, $[t_A,t_A]
\sim t_V$ and $[t_A,t_V] \sim t_A$.  Then by multiplying
eq.(\ref{eq:nonlinear}) on the right by the result obtained after acting
with $R$ and taking the inverse, the quantity $\Sigma \equiv
e^{2i\pi}$ is seen to obey the {\it linear} transformation law
\begin{equation}
\Sigma \to e^{i\epsilon} \, \Sigma \, e^{-i R( \epsilon) } \,. 
\end{equation}
In a more familiar notation, we may write $\epsilon \equiv \epsilon_V
- \epsilon_A$ and $\epsilon_V \equiv \epsilon_V^a t_V^a$, $\epsilon_A
\equiv \epsilon_A^a t_A^a$.  Then
\begin{equation}\label{Sigma}
\Sigma \to e^{i\epsilon_L} \Sigma e^{-i\epsilon_R} \,,
\end{equation}
where $\epsilon_{L,R} \equiv \epsilon_V \mp \epsilon_A$.   
This generalizes eq.(\ref{Utransform}) to the case 
where the elements of $\Sigma$ do not span a full group manifold. 

The choice of variables eq.(\ref{Sigma}) allows us to immediately write down the
topological interactions in the form of a WZW term for models with an 
internal parity operation. 
The chiral current is defined as $\alpha = d\Sigma\, \Sigma^\dagger$, 
and obeys $d\alpha = \alpha^2$. 
The result is simply eq.(\ref{final1}) with $U\to\Sigma$.  
The anomalous gauge variation of the resulting WZW action is 
\begin{align}\label{parityanomaly}
\delta \Gamma_{WZW} 
&= -{N\over 24\pi^2} \int_{M^4} {\rm Tr} \bigg\{ 
(\epsilon_V - \epsilon_A)\left[ (dA_V-dA_A)^2 - {i\over 2}d(A_V-A_A)^3 \right] \nl
&\qquad \qquad \qquad \qquad   - 
(\epsilon_V + \epsilon_A)\left[ (dA_V+dA_A)^2 - {i\over 2}d(A_V+A_A)^3 \right] \bigg\} \,.
\end{align}

As an example, we consider the interactions arising when the Higgs is 
identified with a Nambu Goldstone boson of the symmetry breaking 
$SU(5) \to SO(5)$.  This pattern of symmetry breaking has been incorporated 
into a Little Higgs model by Arkani-Hamed, Cohen, Katz and Nelson~\cite{Arkani-Hamed:2002qy}. 
Let $\Phi$ be a two-index symmetric tensor representation of $SU(5)$, 
developing the VEV 
\begin{equation}\label{omegavev}
\langle \Phi \rangle \equiv \Omega = \left(
\begin{array}{ccccc} 
&&&& 1 \\
&&&1 \\
&&1 \\
&1 \\
1 
\end{array}
\right) \,. 
\end{equation} 
The fourteen NGB's corresponding to 
broken symmetry generators are
\begin{equation}\label{broken}
\pi = \pi^a t_A^a = \left(\begin{array}{ccc} 
\chi^T + \frac12 \eta & H^* & \phi^\dagger \\
H^T & -2\eta & H^\dagger \\
\phi & H & \chi + \frac12 \eta \end{array}\right) \,,
\end{equation}
where $\chi$ is a Hermitian, traceless $2\times 2$ matrix, $\eta$ is a real 
singlet, $H$ is a complex doublet and $\phi$ is a symmetric $2\times 2$ matrix.  
The ten unbroken symmetry generators correspond to 
\begin{equation}\label{unbroken}
A = A^a t^a_V = \left(\begin{array}{ccc} 
-W^T-\frac{g_1}{2} B & C & D\sigma^2 \\
C^\dagger & 0 & -C^T \\
D^*\sigma^2 & -C^* & W + {g_1\over 2} B \end{array}\right) \,, 
\end{equation}
where $W=\sum_{a=1}^3 g W^a\sigma^a/2$ is a Hermitian, traceless
$2\times 2$ matrix, $B$ is a real singlet, $C$ is a complex doublet,
and $D$ is a complex singlet.  The model 
gauges two $SU(2)\times U(1)$ groups, one corresponding to the
standard model $W$ and $B$ in eq.(\ref{unbroken}), and another
corresponding to heavy partners $W^\prime$ and $B^\prime$ that will
eat $\chi$ and $\eta$ in eq.(\ref{broken}).  This implements the notion
of ``collective symmetry breaking'' to stabilize the Higgs mass.

The WZW term can be evaluated straightforwardly.  Its existence is
related to the nontrivial homotopy group $\pi_5(SU(5)/O(5)) = \bm{Z}$.
Since the full WZW term is odd under the internal parity, and
invariant under weak isospin, interactions involving only standard
model fields $W$, $B$, $H$ are forbidden (the parity operation takes
$W,B\to +W,B$ and $H\to-H$).  Interactions do occur involving heavy
partners $W^\prime$, $B^\prime$ of the standard model gauge bosons, or
the Goldstone boson field $\phi$.  For example, interactions involving
the heavy hypercharge field $B^\prime$ are
\begin{multline}\label{result4}
\Gamma_{WZW} \supset {N \over 4\pi^2 F^2 } 
\int_{M^4} (v+h^0)^2 B^\prime\big[ 
g^2(W^+ dW^- + W^- dW^+ + W^3 dW^3) -gg_1( W^3 dB + B dW^3 ) \\
+ g_1^{2} B dB - i g^2 W^+ W^- ( 3g W^3  - g_1 B )  \big]\,.  
\end{multline}
Here $B^\prime$ is written in unitary gauge, having eaten the $\eta$
meson.~\footnote{
Since the $B^\prime$
symmetry is anomalous, other structure, \eg\, leptons, must be present
to cancel this gauge anomaly.}  
Similarly, $W$ and $B$ are written in unitary gauge,
having eaten the Goldstone bosons inside of $H$.  In this gauge 
we write $H \sim (0,v+h^0)^T/\sqrt{2}$.
This WZW term describes ``T parity''-violating interactions, 
\eg\, decays of the single ``T-odd'' field $B^\prime$ into standard model fields.

From eq.(\ref{parityanomaly}), it is straightforward to see that 
the anomalous gauge variation of the WZW action
is precisely that of a fermion theory with $2N$ left-handed fermions transforming 
in the fundamental representation of $SU(5)$:~%
\footnote{
As noted in \cite{Chu:1996fr}, eq.(\ref{parityanomaly}) 
is the gauge variation of a Lagrangian with 
{\it two} sets of Weyl fermions, $\Psi_L$ and $\Psi_R$, 
coupled respectively to the restricted set of gauge fields
$A_V-A_A$ and $A_V+A_A$.  
However, eq.(\ref{parityanomaly}) would describe only a subset 
of the anomalies for such a fermion theory.   
Using trace identities arising from  
the anomaly-free embedding of $SO(5)$ inside of $SU(5)$ 
(\eg, $\Tr[\epsilon_V (dA_V)^2]=0$, $\Tr[\epsilon_A (dA_AdA_V+dA_VdA_A)]=0$, \dots), 
(\ref{parityanomaly}) actually gives precisely 
the nonabelian anomaly for {\it one} set of Weyl fermions. 
} 
\begin{equation}
\Psi_L \to e^{i(\epsilon_V + \epsilon_A)} \Psi_L \,. 
\end{equation}
In a composite theory of underlying fermions, the symmetry breaking corresponds
to a VEV for the operator
\begin{equation}
\epsilon^{\alpha\beta} \psi_{\alpha}^{a i} \psi_{\beta}^{a j} \sim F^3 
\Omega^{ij} \,,
\end{equation}
where $i,j=1..5$ are flavor indices, $\Omega$ is a symmetric matrix as in 
(\ref{omegavev}), 
$a=1..2N$ is a summed color index, and $\alpha,\beta=1..2$ 
are Lorentz indices.   Such a scheme is possible for fermions transforming in a real representation
of the color group---\eg,
in the adjoint representation 
of $SU(N_c)$ for an odd number of colors, $2N=N_c^2-1$, or in the fundamental representation 
of $SO(N_c)$ for an even number of colors, $2N=N_c$.

\section{Summary}

We have summarized the first steps toward the theory
of topological interactions of Higgs bosons. Such
interactions are present when these fields
occur as composite pNGB's,  or more generally in
theories of extra dimensions. 

While the analysis involves novel constructions based on 
topological features of the NGB field manifold, we emphasize that
in the spirit of effective field theory, it is simply a mistake
to omit such interactions. 
As in the familiar case of the QCD chiral Lagrangian, the WZW term 
is a remnant of underlying UV physics, and modifies the predictions of
the low-energy theory. 
For the case of technicolor, 
we found the interactions eq.(\ref{result1}) amongst the pNGB's and gauge bosons. 
For cases of interest to Little Higgs theories, we derived the 
gauged WZW term by two methods, via nonlinear realization of the symmetry group
on a restricted submanifold, or via a direct construction beginning with 
a topological action.  Application to specific models give predictions 
such as eqs.(\ref{nonlinearWZW}) and (\ref{result2}), 
which are analogs of $\pi^0\to\gamma\gamma$.   These 
interactions directly probe the structure of the underlying fermion
UV completion theory, \eg, 
allowing one to count the number of colors 
in an underlying strong gauge group.   

We also pointed out the simple structure of the gauged WZW term for 
symmetry-breaking patterns that respect an internal parity.  
This includes the case of the QCD chiral Lagrangian, and the case of 
Little Higgs models with $T$ parity.  If such models arise from composite 
fermions, we see that $T$ parity cannot be an exact symmetry.  
As an example, eq.(\ref{result4}) is 
an interaction between the single ``$T$ odd'' partner of the 
hypercharge gauge boson and ``$T$ even'' standard model particles.  

Much work remains in order to explore the phenomenological implications 
of WZW interactions for Little Higgs bosons, 
and to identify the most promising signatures 
at future colliders such as the LHC, ILC, and even beyond to 
CLIC or a muon collider. 

\vskip .1in
\noindent
{\bf Acknowledgements}
\vskip .1in
We thank W. Bardeen for helpful discussions. 
This work was hosted at Fermilab 
in the Fermilab Theoretical Physics Department
operated by Fermi Research Alliance, LLC  under Contract No. 
DE-AC02-07CH11359 with the United States Department of Energy.



\begin{thebibliography}{99} 

\bibitem{Kaplan:1983fs}
  D.~B.~Kaplan and H.~Georgi,
  Phys.\ Lett.\ B {\bf 136}, 183 (1984);\\
  D.~B.~Kaplan, H.~Georgi and S.~Dimopoulos,
  Phys.\ Lett.\ B {\bf 136}, 187 (1984);\\
  H.~Georgi, D.~B.~Kaplan and P.~Galison,
  Phys.\ Lett.\ B {\bf 143}, 152 (1984);\\
  H.~Georgi and D.~B.~Kaplan,
  Phys.\ Lett.\ B {\bf 145}, 216 (1984);\\
  M.~J.~Dugan, H.~Georgi and D.~B.~Kaplan,
  Nucl.\ Phys.\ B {\bf 254}, 299 (1985).
  
\bibitem{nima}
  N.~Arkani-Hamed, A.~G.~Cohen and H.~Georgi,
  Phys.\ Lett.\ B {\bf 513}, 232 (2001)
  [hep-ph/0105239];\\
    N.~Arkani-Hamed, A.~G.~Cohen, E.~Katz, A.~E.~Nelson, T.~Gregoire and J.~G.~Wacker,
  JHEP {\bf 0208}, 021 (2002)
  [hep-ph/0206020].

\bibitem{Wess:1971yu}
  J.~Wess and B.~Zumino,
  Phys.\ Lett.\ B {\bf 37}, 95 (1971).

\bibitem{Witten}
  E.~Witten,
  Nucl.\ Phys.\ B {\bf 223}, 422 (1983).


\bibitem{BJ} 
  J.~S.~Bell and R.~Jackiw,
  Nuovo Cim.\ A {\bf 60}, 47 (1969).

\bibitem{Adler}
  S.~L.~Adler,
  Phys.\ Rev.\  {\bf 177}, 2426 (1969).
  
\bibitem{bardeen}
  W.~A.~Bardeen,
  Phys.\ Rev.\  {\bf 184}, 1848 (1969).
  
\bibitem{kay}
  O.~Kaymakcalan, S.~Rajeev and J.~Schechter,
  Phys.\ Rev.\ D {\bf 30}, 594 (1984).

\bibitem{man}
A.~Manohar and G.~W.~Moore,
Nucl.\ Phys.\ B {\bf 243}, 55 (1984).

\bibitem{tye}
  H.~Kawai and S.~H.~H.~Tye,
  Phys.\ Lett.\ B {\bf 140}, 403 (1984).

\bibitem{hillch}
  C.~T.~Hill,
  Phys.\ Rev.\ D {\bf 73}, 126009 (2006)
  [hep-th/0603060];\\
   C.~T.~Hill,
  Phys.\ Rev.\ D {\bf 73}, 085001 (2006)
  [hep-th/0601154];\\
  C.~T.~Hill and C.~K.~Zachos,
  Phys.\ Rev.\ D {\bf 71}, 046002 (2005)
  [hep-th/0411157].
  
\bibitem{Kaplan:2003uc}
  D.~E.~Kaplan and M.~Schmaltz,
  JHEP {\bf 0310}, 039 (2003)
  [hep-ph/0302049].

\bibitem{2hill}
  C.~T.~Hill and R.~J.~Hill, in preparation. 

\bibitem{Cheng:2003ju}
  H.~C.~Cheng and I.~Low,
  JHEP {\bf 0309}, 051 (2003)
  [hep-ph/0308199].

\bibitem{Arkani-Hamed:2002qy}
  N.~Arkani-Hamed, A.~G.~Cohen, E.~Katz and A.~E.~Nelson,
  JHEP {\bf 0207}, 034 (2002)
  [hep-ph/0206021].

\bibitem{AdlerBardeen}
  S.~L.~Adler and W.~A.~Bardeen,
  Phys.\ Rev.\  {\bf 182}, 1517 (1969).

\bibitem{Coleman:1969sm}
  S.~R.~Coleman, J.~Wess and B.~Zumino,
  Phys.\ Rev.\  {\bf 177}, 2239 (1969);\\
   C.~G.~.~Callan, S.~R.~Coleman, J.~Wess and B.~Zumino,
  Phys.\ Rev.\  {\bf 177}, 2247 (1969).

\bibitem{Gasser:1984gg}
  J.~Gasser and H.~Leutwyler,
  Nucl.\ Phys.\ B {\bf 250}, 465 (1985);\\
   A.~J.~Buras, A.~Poschenrieder, S.~Uhlig and W.~A.~Bardeen,
  hep-ph/0607189.

\bibitem{Witten:1982fp}
  E.~Witten,
  Phys.\ Lett.\ B {\bf 117}, 324 (1982).

\bibitem{D'Hoker:1994ti}
  E.~D'Hoker and S.~Weinberg,
  Phys.\ Rev.\ D {\bf 50}, 6050 (1994)
  [hep-ph/9409402].

\bibitem{Hull:1990ms}
  C.~M.~Hull and B.~J.~Spence,
  Nucl.\ Phys.\ B {\bf 353}, 379 (1991).

\bibitem{Chu:1996fr}
  C.~S.~Chu, P.~M.~Ho and B.~Zumino,
  Nucl.\ Phys.\ B {\bf 475}, 484 (1996)
  [hep-th/9602093].

\end{thebibliography}
\end{document}